\documentclass[preprint,aps,pra,showpacs]{revtex4}
\usepackage{amssymb}

\usepackage{graphicx}

\begin{document}
\title{ Quantum entanglement via two-qubit quantum Zeno dynamics}
\author{Xiang-Bin Wang}\affiliation{Frontier Research System, The Institute of Physical
and Chemical Research (RIKEN), Wako-shi 351-0198, Japan}
\affiliation{CREST and Imai QCI project of ERATO-SORST, Japan
Science and Technology Agency (JST), Kawaguchi, Satima 332-0012,
Japan} \affiliation{Department of Physics, Tsinghua University,
Beijing 100084, China}
\author{J. Q. You}
 \affiliation{Advanced Science Institute, The Institute of
Physical and Chemical Research (RIKEN), Wako-shi 351-0198,
Japan}\affiliation{Department of Physics and Surface Physics
Laboratory (National Key Laboratory), Fudan University, Shanghai
200433, China}
 \author{ Franco Nori}\affiliation{Frontier Research System, The Institute of Physical
and Chemical Research (RIKEN), Wako-shi 351-0198, Japan}
\affiliation{Center for Theoretical Physics, Physics Department,
Center for the Study of Complex Systems, The University of
Michigan, Ann Arbor, MI 48109-1040, USA}
\begin{abstract}
We study the two-particle quantum Zeno dynamics with a type of
non-deterministic collective measurement whose outcome indicates
whether the two-particle state has been collapsed to $|11\rangle$.
Such a threshold detection, when used continuously, can lead to
non-trivial quantum dynamics. We show that such type of dynamics can
be used to produce quantum entanglement almost deterministically.
We then numerically show the robustness of the method and we find
that the operational errors of the small-angle rotations do not
accumulate. We also propose a possible implementation  using
superconducting flux qubits.
\end{abstract}
\maketitle
\section{ Introduction}
Due to  the quantum Zeno effect \cite{zeno1,zeno0,zeno},  quantum
decay can be suppressed if a particle is continuously observed in
the same basis. However, if we continuously observe a particle with
a slowly changing basis, the particle's state will keep following
the measurement basis. As we shall show, applying this fact to a
multi-qubit system, one can make non-trivial quantum-state steering.
The quantum Zeno effect of a {\it single} particle has attracted
considerable interest in the past. Recently, the {\em
multi}-particle dynamics due to the quantum Zeno effect has also
been studied \cite{franson,multi}.  In particular, Franson's group
proposed to realize the controlled-NOT gate in quantum computing
through the quantum Zeno effects of two optical qubits. Here we
study a {\it two}-qubit quantum Zeno dynamics with threshold
detection, which checks whether the two-particle state has been
collapsed to $|11\rangle$. As we shall show below, such measurements
can suppress the coefficient of the state $|11\rangle$ and can lead
to non-trivial two-particle and multi-particle entangled states,
e.g., Bell states, GHZ states, cluster states, and so on.

 Quantum entanglement is an important resource for quantum
information processing (QIP). Two-qubit joint operations are crucial
for tasks such as creating and manipulating quantum entanglement in
QIP. Indeed, controlled-NOT (CNOT) gates and single-qubit unitary
transformations are sufficient for generating any quantum
entanglement and for universal quantum computing. However,
implementing a CNOT gate experimentally seems to be a daunting task.
This is a huge barrier to scalable quantum computing, which requires
numerous CNOT gates. To avoid this difficulty, it has been
proposed~\cite{knill,gottc,pittman0,nil,others} to replace the CNOT
gate with a Bell measurement . Indeed, it is possible to replace
CNOT gates by quantum teleportation \cite{gottc}, where the only
collective operation is a Bell measurement. However, so far it is
unknown how to do a projective Bell measurement without using a CNOT
gate. In other words, the complete projective Bell measurement seems
to be as difficult to implement experimentally as a CNOT gate.
 In practice, a non-deterministic collective measurement is often
used because it is easier to implement. However, many of these
proposals  can only realize a probabilistic QIP.

An elegant alternative, one-way quantum computation using cluster
states \cite{cluster1}, is promising. In that approach, cluster
states are first produced and afterwards used for quantum computing
through individual measurements only.  Efforts have been made
towards the efficient generation of cluster states via
non-deterministic two-qubit measurements. Also, there are proposals
for generating cluster states using solid-state qubits (see, e.g.
\cite{Loss,Wein,Tana,YWTN,Xue}).

Non-deterministic collective measurements, as already demonstrated
in a number of experiments, can be used to produce entangled states
including cluster states probabilistically. Indeed, cluster-state
quantum computation has recently been demonstrated with such a
technique \cite{zei}.

Here we present an alternative approach. We show that one can
actually produce, almost {\em deterministically},  quantum
entanglement, such as a cluster state via non-deterministic
measurements, we name  ``threshold measurements" or
``$J-$measurements". These indicate whether the measured two-qubit
state is $|11\rangle$. Although the measurement outcome itself is
non-deterministic, by using the quantum Zeno effect (see, e.g.,
\cite{zeno1,zeno0,zeno}),  a certain quantum subspace is almost
inhibited from decay if it is measured continuously; therefore
providing an almost deterministic result.

Consider the following two-qubit non-deterministic collective
measurement composed of two projectors:
\begin{equation}
J_1=|1\rangle\langle 1|\otimes |1\rangle\langle 1|,~~ J_0={\cal
I}-J_1\, ,
\end{equation}
where ${\cal I}$ is the four-dimensional identity operator. We call
this type of measurement  ``$J-$measurement''. This $J-$measurement
is different from a parity measurement or singlet-triplet
measurement \cite{pittman0,rudolph,kwek}. Our measurement is a {\em
threshold measurement} on whether both qubits are in state
$|1\rangle$.

As shown below, technically, a $J-$measurement does not need to
control exactly the interaction strength or duration, neither does
it assume any synchronization difficulty. One only needs to turn on
the ``threshold detector'' to see whether the current is larger than
the threshold value.  For clarity, we assume that a two-qubit state
is monitored by a $J$-measurement detector: If the detector clicks,
the state is collapsed to $|11\rangle$; if the detector does not
click, the state is projected into the subspace
\begin{equation}
J_0 =\{|00\rangle,|01\rangle,|10\rangle\}\;.
\end{equation}
If the state is initially in the subspace $J_0$, the quantum Zeno
effect will inhibit the state to evolve to $|11\rangle$ if the $J$
measurement is performed frequently. We shall show that, by only
using {\em single}-qubit operations and $J$-measurements, one can
almost  deterministically produce  large cluster states {\em
without} using any other separate conditional dynamics or quantum
entangler.

Our work is related to prior works as mentioned already, in
particular, the one by Franson's group \cite{franson}: both
Ref.\cite{franson} and our work propose to apply the two-qubit
quantum Zeno for quantum information processing (QIP). However, they
are different in many aspects:  1) The roles of the quantum Zeno
effects in these two works are different. In Franson's design, one
needs both a quantum entangler (a coupled optical fiber or a beam
splitter) and the quantum Zeno effect (through a two-photon
absorbtion by an atomic gas). In our design, we {\em only} need a
threshold measurement and a single-qubit rotation. Except for these,
we need neither a separate quantum entangler nor a two-qubit quantum
unitary. In other words, in our design, the quantum Zeno effect has
a more crucial role: it {\em produces} the quantum entanglement
rather than assisting the separate quantum entangler for QIP. 2) The
calculations and the results are different. In our design, since we
only use the threshold measurement, the dynamics due to the quantum
Zeno effect is more complicated. The effects of the errors caused by
discrete measurements are studied in details in our work. (These
results are not limited by any specific physical systems.)  3) The
proposed physical systems for experimental realization are
different. Franson's group studies the optical system with
two-photon absorbtion by atoms while here we consider solid-state
qubits with our threshold detection. Different systems  have
different advantages in various aspects, such as technical overhead,
cost, robustness, scalability, and so on. Therefore, studies of
different physical systems are needed.

\section{ Quantum entanglement through a quantum Zeno effect based on
$J-$measurements}

Let us define
$$|\psi^{\pm}\rangle=\frac{1}{\sqrt
2}(|01\rangle\pm|10\rangle)\;.$$
We now show how to drive the
two-qubit state $|00\rangle$ to the maximally entangled state
$|\psi^+\rangle$ by repeating the following $W$ operation (on the
two qubits): \\({\em i})~Rotate each individual qubit by the same
small angle $\theta$, and then  \\({\em ii})~perform a
$J-$measurement.

After a number of $W$ operations, the state $|00\rangle$ can be
driven into $|\psi^+\rangle$ with probability $1-O(\sin\theta)$. We
do not have to require a constant $\theta$ for each application of
$W$, but to simplify the presentation we now assume a constant
positive $\theta$ for each step. The initial state is
\begin{eqnarray}
|\chi_0\rangle \;\!&\!=\!&\!\; |00\rangle\;=\; a_0|00\rangle + \sqrt
2b_0\;|\psi^+\rangle \nonumber\\
\!&\!=\!&\!\;a_0|00\rangle+b_0(|01\rangle+|10\rangle)\;,
\end{eqnarray}
with $a_0=1$ and $b_0=0$. After the first $W$ operation, the initial state becomes
\begin{equation}
|\chi_1\rangle = a_1|00\rangle + b_1 (|01\rangle+|10\rangle)\;,
\end{equation}
with probability \begin{equation}N_1 =1-\sin^4\theta \approx
1\;.\end{equation} Here $$a_1=\cos^2\theta/N_1\;,$$ and
$$b_1=\sin\theta\cos\theta/N_1\;.$$ Thus, the probability amplitude of
$|\psi^+\rangle$ increases after each step. Through the iterative
application of $W$, the state $|\chi_0\rangle$ will, sooner or
later, be projected into $|\psi^+\rangle$. Therefore, we only need
to show that after less than $k_1\sim O(1/\sin\theta)$ applications
of $W$, the two-qubit quantum state $|\chi_0\rangle$ is mapped into
$|\psi^+\rangle$ with high probability. In this case, the total
probability that the state $|\chi_0\rangle$ is projected into
$|11\rangle$ during the whole process is only $O(\sin\theta)$.
Therefore, given a sufficiently small $\theta$, the failure
probability is negligible and the result is almost deterministic.

Let us consider now the state $|\chi_i\rangle$ obtained after $W$ is
applied $i$ times to $|\chi_0\rangle$:
\begin{equation}
|\chi_i\rangle =W^i|\chi_0\rangle=a_i|00\rangle +
b_i(|01\rangle+|10\rangle)\;.
\end{equation}
Assume that $a_i,b_i\ge 0$. After applying $W$ one more time we
obtain
\begin{equation}
|\chi_{i+1}\rangle =W|\chi_i\rangle=a_{i+1}|00\rangle +
b_{i+1}(|01\rangle+|10\rangle)
\end{equation}
with $$a_{i+1}=
[a_i(1-\sin^2\theta)-2b_i\sin\theta\cos\theta]/N_{i+1}\;;$$ $$
b_{i+1}=[b_i(1-\sin^2\theta) + a_i \sin\theta\cos\theta]/N_{i+1}$$
with
$$N_{i+1}=1-a_i^2\sin^4\theta-4b_i^2\sin^2\theta\cos^2\theta\sim
1-O(\sin^2\theta)\;.$$ The amplitude difference between $|00\rangle$
and $|\psi^+\rangle$ changes after each step. We define
\begin{eqnarray}
\delta_{i+1} \!&\!=\!&\! b_{i+1}-a_{i+1}-(b_{i}-a_i) \nonumber\\
\!&\!=\!&\! \left[(a_i+2b_i)\sin\theta\cos\theta
+(a_i-b_i)\sin^2\theta \right]/N_{i+1}\;.\nonumber\end{eqnarray}
After $k_1$ applications of $W$, we obtain
$$b_{k_1}-a_{k_1} = b_0-a_0+\sum_{i=1}^{k_1}\delta_i.$$
Our goal now is to know how large $k_1$ must be so that $$a_{k_1}\sim 0,$$ i.e.,
$$(b_{k_1}-a_{k_1})\sim1/\sqrt 2\;.$$ If all $\{a_i,b_i;~i\le k_1\}$ are
non-negative, then $$\delta_{i+1}\ge \sin\theta \cos\theta \;,$$
therefore
\begin{equation}
b_{k_1}-a_{k_1} \ge -1 +k_1 \sin\theta\cos\theta\;.
\end{equation}
Given this, we conclude that there exists a positive number
\begin{equation}k_1\sim O\left(1/\sin\theta\right)\;,\end{equation}
  such that after $W$ is
applied $k_1$ times, $a_{k_1}$ must be almost zero, provided that
$\theta$ is sufficiently small. From the above derivation and a
similar derivation, we draw the following lemma:

{\bf Lemma:} {\em Iterating the  $W$
operation can map the state $|00\rangle$ into $|\psi^{+}\rangle$,
and also map the state $|\psi^+\rangle$ into $-|00\rangle$ in the
same number of steps. Together with single-qubit unitary operations,
any state $\alpha|00\rangle+\beta |\psi^+\rangle$ can be mapped into
$|\psi^+\rangle$ with less than $k_1=O(1/\sin\theta)$ iterations of
$W$.}

Iterating the $W$ operation can also map the initial state
$|10\rangle$ into the maximally entangled state $|\psi^+\rangle$.
This can be seen as follows: Consider now the initial state
\begin{equation}
|\chi_0'\rangle =|10\rangle =\frac{1}{\sqrt
2}(|\psi^+\rangle-|\psi^-\rangle)\;.
\end{equation}
The state $|\psi^-\rangle$ is invariant under identical individual
rotations. Also, $|\psi^+\rangle$ can be mapped into $-|00\rangle$
(see the Lemma above). Therefore, we obtain the state
\begin{equation}|\chi'_{k_1}\rangle~\thickapprox~-\frac{1}{\sqrt
2}(|00\rangle+|\psi^-\rangle)\end{equation} after $k_1$ iterations
of $W$. After applying a local phase-flip, the state is changed into
\begin{equation}
|\chi'\rangle~\thickapprox~\frac{1}{\sqrt
2}(|00\rangle+|\psi^+\rangle).
\end{equation}
Again using our Lemma above we conclude that this state can also
be mapped into $|\psi^+\rangle$.

\section{ Quantum dynamics of the {\em W} operator}

We now study more
precisely the properties of $W$  using its matrix representation.
Given any initial state $|\gamma\rangle$, after a $W$ operation, the
(un-normalized) state in the $J_0$ space becomes:
\begin{equation}
|\gamma_1\rangle =M(\theta)|\gamma\rangle=J_0 \,R(\theta)\otimes
R(\theta)|\gamma\rangle
\end{equation}
where $M(\theta)$ is the matrix representation of $W$. The
probability that the qubit is projected into the $J_0$ subspace is
$|\langle\gamma_1|\gamma_1\rangle|^2$. In  matrix representation,
$$R(\theta)= \left(\begin{array}{cc}
\cos\theta &-\sin\theta\\
\sin\theta & \cos\theta
\end{array}\right),
~~~|0\rangle=\left(\begin{array}{c}1\\0\end{array}\right),
~~~|1\rangle=\left(\begin{array}{c}0\\1\end{array}\right),$$
and $J_0=I_3\oplus 0$ ($I_3$ is the $3\times 3$ identity matrix).
Since we are only interested in the case when the initial state
$\gamma\in J_0$, the matrix representation for a $W$ operation in
$J_0$ space is simplified to
\begin{eqnarray}\label{evolution}
M(\theta)= \left(\begin{array}{ccc}\cos^2\theta &
-\sin\theta\cos\theta & -\sin\theta\cos\theta
\\\sin\theta\cos\theta  &\cos^2\theta
&-\sin^2\theta
\\\sin\theta\cos\theta & -\sin^2\theta &\cos^2\theta
\end{array}\right).
\end{eqnarray}
In this matrix representation, the ket states are represented by
\begin{eqnarray}
[|00\rangle, |10\rangle, |01\rangle ]=[\left(\begin{array}{c}
1\\0\\0\end{array}\right), \left(\begin{array}{c}
0\\1\\0\end{array}\right),\left(\begin{array}{c}
0\\0\\1\end{array}\right)].
\end{eqnarray}
Hereafter $M(\theta)$ is simply denoted by $M$.  After $N$
iterations of $W$, the evolution operator in the $J_0$ subspace is
 $M^N$. We now test our results numerically. First, we
iterate $W$ for $k_1=100$ times with $\theta={\pi}/({200\sqrt 2})$.
We then obtain the numerical matrix \begin{eqnarray} M^{100}=\left(
\begin{array}{ccc}
0.0039& -0.7028 &-0.7028\\ 0.7028 & 0.4980 & -0.5020\\
0.7028 & -0.5020 & 0.4980
\end{array}
\right)\;.\label{m100}\end{eqnarray} This shows that if we start
from the initial state $|00\rangle$, after 100 iterations of $W$, we
obtain the maximally entangled state $|\psi^+\rangle$ with
probability $98.8\%$ and a fidelity  larger than $99.99\%$.
Iterating $W$ 1000 times with $\theta=\pi/(2000\sqrt 2)$, we obtain
a highly entangled state: with 99.9\% probability and a fidelity
larger than $1-10^{-6}$.

\section{Intelligent evolution}

If the initial state is $|10\rangle$, after iterating the {\em W}
operator, we can also obtain the maximally entangled state
$|\psi^-\rangle$. Also, we want to have an ``intelligently-designed"
evolution which will produce different maximally entangled states
depending on whether the initial state is $|00\rangle$ or
$|10\rangle$, since this type of evolution is crucial in expanding a
cluster state, as shown below. After $k_1=100$  iterations of $W$,
we perform a phase flip operation
$P=\left(\begin{array}{cc}1&0\\0&-1\end{array}\right)$ to the first
qubit, and apply the $W$ operation $k_2=50$ times to obtain the
final evolution matrix
\begin{eqnarray}\label{mpm}
M^{50}PM^{100}= \left(
\begin{array}{ccc}
0.0027& 0.0011 &-0.9958\\ -0.7008 & -0.6994 & 0.0027\\
0.7047 & -0.7033 & -0.0012
\end{array}
\right)\;.\label{mp}
\end{eqnarray}
As shown below, such an ``intelligent" evolution can expand a
cluster state deterministically. In the above three-stage
operations, $W$ was iterated $k_1$ times, then a phase
flip $P$ was applied, and finally $k_2$ iterations of $W$. If $\theta$ is
{\em very} small, the constraints $k_1\theta ={\pi}/({2\sqrt 2})$ and
$k_2\theta ={\pi}/({4\sqrt 2})$ will produce almost perfect results
(i.e., with both the probability and the fidelity  almost equal to
1). Now we show this explicitly. Suppose that after $k_1$ iterations
of $W$, the initial state $|00\rangle$ is mapped into the maximally
entangled state $|\psi^+\rangle$. This requires $m_{11}$ (the matrix
element of the first row and the first column of the matrix
$M^{k_1}$) to be exactly 0. Here, \begin{equation}M^{k_1} =
\left[\cos^2\theta I_3 + r(\theta)\right]^{k_1} ,\end{equation}and
\begin{eqnarray}r(\theta)=\left(\begin{array}{ccc} 0 &
-\sin\theta\cos\theta
&-\sin\theta\cos\theta \\
\sin\theta\cos\theta&0&-\sin^2\theta\\
\sin\theta\cos\theta & -\sin^2\theta&0
\end{array}\right).\end{eqnarray}
Therefore
\begin{equation}\label{tt1}
M^{k_1}=\sum_{n=0}^{k_1}C_{k_1}^n \;r^n\cos^{2k_1-2n}\theta,
\end{equation}
with
$C_{k_1}^n=\left(\begin{array}{c}n\\k_1\end{array}\right)={k_1(k_1-1)\cdots
(k_1-n+1)}/{n!}$\,. Any term of the form  $k_1^l\sin^j\theta$ is
discarded in the summation  if $j>l$ because $\theta$ is very small.
Therefore, we obtain
\begin{equation}\label{cos}
m_{11} ~\approx ~\cos (\sqrt 2 k_1\sin\theta)
\end{equation}
which becomes 0 when \begin{equation}k_1\theta\;
=\;\frac{\pi}{2\sqrt 2}.\end{equation} Consider now another initial
state
\begin{equation}|10\rangle =(|\psi^+\rangle-|\psi^-\rangle)/{\sqrt 2}.\end{equation} The
$|\psi^-\rangle$ part is invariant under $W$. According to our lemma
above, after $k_1$ iterations of $W$, the state $|10\rangle$ must be
changed to \begin{equation}(|\psi^-\rangle-|00\rangle)/{\sqrt 2}
.\end{equation} After the phase-flip $P$ is applied, the state
becomes \begin{equation}|\chi\rangle\;=\;-
(|\psi^+\rangle+|00\rangle)/{\sqrt 2}.\end{equation}
Let us recall now the evolution property for the initial state $|00\rangle$ under
iterations of $W$. According to our Lemma, after $k_2$ iterations of
$W$ with $k_2\theta\;=\;\pi/(4\sqrt 2)$, the state $|\chi\rangle$
becomes $-|\psi^+\rangle$. This means, if we start from
$|\chi\rangle$, we only need
\begin{equation}k_2\theta\;=\;\frac{\pi}{4\sqrt 2}\end{equation} in order to obtain
$-|\psi^{+}\rangle$. Based on these facts we conclude the following theorem:

{\bf Theorem:} {\em The operator $W^{k/2}\,P\,W^{k}$ can change
 the initial states $(|00\rangle,~|10\rangle)$ into
$(|\psi^-\rangle,~-|\psi^+\rangle)$ if $k\theta=\pi/(2\sqrt 2)$, and
the $\theta$ for  every step is very small.}

Our $W$ operation is not limited to produce two-qubit entanglement,
as shown below; it can also be used to {\em expand} a cluster state
almost deterministically.

\section{Quantum entanglement expansion}

As is known~\cite{cluster1}, one can
build a large cluster state from the product state
$|+\rangle|+\rangle\cdots|+\rangle$, where
$|+\rangle=(|0\rangle+|1\rangle)/\sqrt 2$, with a controlled-phase
(C-Phase) gate applied to the nearest qubits from the left to the
right. A C-Phase will change any state $|i\rangle|j\rangle$ into
$(-1)^{ij}|i\rangle|j\rangle$, ($i,j \in 0,1$). For example,
consider the two-qubit case: The state $|+\rangle|+\rangle$ is
changed into $(|0\rangle|+\rangle+|1\rangle|-\rangle)/\sqrt 2$,
which can be transformed into $|\psi^+\rangle$ by a single-qubit
flip operation. In general, an $n$-qubit cluster state can be
written in the following bipartite form
\begin{equation}\label{e14}
|C_n\rangle ~= ~|E\rangle|0\rangle \,+\,|E'\rangle|1\rangle
\end{equation}
where $|E\rangle$ and $|E'\rangle$ span the subspace of the first
$(n-1)$ qubits, $|0\rangle$ and $|1\rangle$ span the subspace of
the $n${th} qubit. We can expand this to an $(n+1)$-qubit cluster
state using a C-Phase gate with an ancilla qubit $|+\rangle$.
Explicitly, after the C-Phase gate, the expanded cluster state
becomes
\begin{equation}\label{np1}
|C_{n+1}\rangle \,=\,|E\rangle|0\rangle|+\rangle \,+\,
|E'\rangle|1\rangle|-\rangle.
\end{equation}
The few lines above are  known results on how to produce a cluster
state with C-Phase gates~\cite{cluster1}. Below we show how to
expand a cluster state in the form of Eq.~(\ref{np1}) by our $W$
operations. Here, we do not need any C-Phase gate since the $W$
operation is sufficient for such type of expansion. We first take a
Hadamard transform of the last qubit of the initial $n$-qubit
cluster state in Eq.~(\ref{e14}) and we set the ancilla state to be
$|0\rangle$. The entire state of the $(n+1)$ qubits is now
\begin{eqnarray}\label{dd}
|D\rangle \!&\!=\!&\! \left(|E\rangle|+\rangle\,+\,
|E'\rangle|-\rangle\right)\otimes |0\rangle\nonumber\\
\!&\!=\!&\!\frac{1}{\sqrt
2}\left[|E\rangle(|00\rangle+|10\rangle)+|E'\rangle(|00\rangle-|10\rangle)\right].
\end{eqnarray}
According to our theorem, the operator $W^{k/2}\,P\,W^{k}$ leads
to the following transformation
\begin{equation}
|00\rangle\longrightarrow
|\psi^-\rangle;~~|10\rangle\longrightarrow -|\psi^+\rangle
\end{equation}
if $k\theta =\pi/(2\sqrt 2)$.  This means that, after applying
$W^{k/2}PW^{k}$, the state of $(n+1)$ qubits becomes
\begin{eqnarray}
&&\frac{1}{\sqrt 2}\left[|E\rangle\left(|\psi^-\rangle
-|\psi^+\rangle\right) + |E'\rangle\left(|\psi^-\rangle
+|\psi^+\rangle \right)\right]\nonumber\\
&&=-|E\rangle|10\rangle \,+\, |E'\rangle|01\rangle .
\end{eqnarray}
After applying a phase-flip and a Hadamard transform to the last two
qubits, and a bit-flip to the $n$th qubit, we obtain an
$(n+1)$-qubit entangled state identical to that of Eq.~(\ref{np1}).
This means that the $J-$measurement can be used to produce and
expand a cluster state almost deterministically, if the rotation
angle $\theta$ of every step is sufficiently small.

\section{Robustness analysis}

In practice, any protocol  always has errors. In our protocol, there
are many iterations. Since our scheme measures the qubits
frequently, one natural question raised here is: If there are small
errors in each iteration, will these errors accumulate and finally
lead to the failure of this scheme? Here we make a partial
investigation of this problem. In each step, we need to rotate {\em
both} qubits by a small angle $\theta$. Intuitively, there could be
operational errors in doing the rotation. Say, sometimes the rotated
angle is larger than $\theta$, and sometimes it is smaller than
$\theta$. Here we do numerical simulations to determine the final
effects of such operational errors with two assumptions: 1) In each
step, the rotated angles of each qubits are the same. 2) There are
only occasional errors in the rotation. Say, at step $i$, the
rotation angle can be $\theta_i=\theta+\epsilon_i$ which is
different from $\theta$, but each $\epsilon_i$ is random.

To have a quantitative evaluation of the robustness of our protocol,
we define $P_s$ as the probability of  obtaining a perfect result,
averaged over the results from the initial states of $|00\rangle$
and $|01\rangle$. Explicitly
\begin{equation}\label{ps}
P_s=\frac{1}{2}\left(|\langle\psi^-
|W^{k/2}\,P\,W^{k}|00\rangle|^2+|\langle\psi^+
|W^{k/2}\,P\,W^{k}|10\rangle|^2\right)
\end{equation}
Here we have taken into account both the probability that the
measurement outcome goes beyond the $J_0$ subspace and the
probability that the final state is not $|\psi^+\rangle$, although
the outcome is in the subspace $J_0$.  Contrary to one's intuition,
the more iterations of $W$ are taken, the less the outcome state is
affected by the operational errors, as shown by the numerical test
in Table I. From the numerical results there we can see that even
for a not-so-large-number of steps, e.g., $k=50$, fairly good
results can be obtained under quite large operational errors
($50\%$).
As shown in the table, in the case when the largest error in
every step is bounded by $50\%$,
the average fidelity is larger than
$96\%$.

\begin{table}
\caption{\label{tab:table1}Numerical results of the $P_s$ values as
defined in Eq.(\ref{ps}) given different operational errors. These
numerically test the robustness of our results with respect to
random operational errors in the rotation. Here $\epsilon_M$ is the
largest possible error of the rotation angle in every step  (in
percentage), $k$ indicates the number of iterations of the operator
$W^{k/2}\,P\,W^{k}$.}
\begin{ruledtabular}
\begin{tabular}{cccccc}
 $~~~~~~~~~~~~~~k~~\setminus~~\epsilon_M\rightarrow$ & $0\%$&$5\%$ & $10\%$ & $20\%$ & $50\%$ \\
\hline
 $50$ & $97.17\%$&  $97.18\%$ & $97.13\%$ & $97.07\%$ & $96.69\%$\\
 $100$ & $98.58\%$&  $98.58\%$ & $98.58\%$ & $98.57\%$ & $98.50\%$\\
 $1000$ & $99.87\%$&  $99.86\%$ & $99.86\%$ & $99.85\%$ & $99.85\%$\\
\end{tabular}
\end{ruledtabular}
\end{table}

Above we have presented our results on the two-qubit quantum Zeno effect
and its application in generating and expanding quantum
entanglement. We find that good fidelity can be achieved even if we
only use fewer than 100 steps with operational errors (occasional
error) up to $50\%$ in every step. The final question remaining is
how to physically implement the $J-$measurement, which is a two-qubit
threshold measurement.

\section{Implementation}

There have been a number of  proposals~\cite{science,PRB,ruskov} for
two-qubit measurements based on quantum dots or superconducting
qubits. There~\cite{science,PRB,ruskov}, not only two-qubit
measurement schemes are given, but also their feasibility, including
decoherence. On the other hand, threshold detections
for a single-qubit have been experimentally demonstrated
already~\cite{science1,science2}.

Compared with the
one-qubit threshold measurement, the two-qubit threshold measurement
does not need extra precise control of interaction or
synchronization.  Here  we consider an implementation scheme for
two-qubit threshold detection, using Josephson-junction  circuits (see, e.g.,
\cite{science1,science2,today,you0,majer}).

Consider a circuit with one large junction, denoted by ``0" and two
parallel flux qubits, each one consisting of three smaller
junctions, as shown in Fig. 1. If the current across junction 0 is
larger than a certain critical value $I_{T0}$, it switches from the
superconducting state to the normal state. The direction of the
current contributed by any qubit in the circuit depends on its
state, say, $|1\rangle$ for the ``up" current and $|0\rangle$ for
the ``down" one. The current contributed from those three-junction
flux qubits  is significantly less than $I_{T0}$. However, with an
appropriate bias current, the current contributed by those flux
qubits determines whether the large junction, 0, will be switched to
the non-superconducting state with a nonzero voltage $V$. The
current is determined by the quantum state of those flux qubits in
the circuit. Suppose that the state $|1\rangle, |0\rangle$ of each
individual qubit contributes a current $\pm { I}_D$, respectively.
If the bias current is set to be, e.g., $I_b=I_{T0}-I_D$, by
monitoring the voltage $V$, we can conclude whether the state of
those flux qubits has been projected to the state $|11\rangle$.
\begin{figure}
\includegraphics[width=3.0in,
bbllx=8,bblly=449,bburx=458,bbury=676]{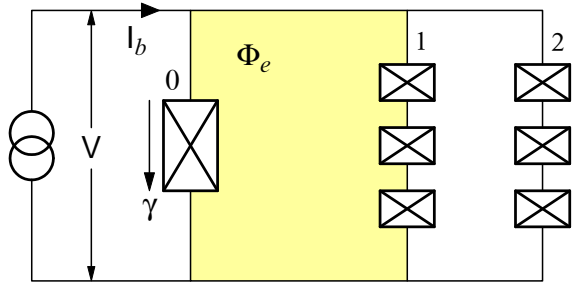}
\caption{\label{Fig1} (Color online) The so called $J-$measurement
can be implemented by a Josephson-junction circuit with flux qubits.
Junction ``0" is a larger junction. Flux qubit 1 and flux qubit 2
each consists of three small junctions. $\Phi_e$ is the flux of the
external magnetic field threading the loop connecting junction ``0"
and qubit 1. }
\end{figure}
Of course, the bias current  $I_b$ and the magnetic flux $\Phi_{e}$ can be
tuned.
Consider the case where there are only two qubits. There are two
subspaces, $J_0=\{|00\rangle,|01\rangle,|10\rangle\}$ and $
J_1=|11\rangle$. A state in subspace $J_1$ ($J_0$) will cause (not
cause) junction ``0" to switch from the superconducting to the
normal state, given a certain bias current $I_b$ and an external
field $\Phi_e$. Thus, when the current $I_b$ is biased, we can
conclude whether the quantum state of those observed qubits
belongs to subspace $J_0$ or  $ J_1$, by monitoring the voltage
$V$.  If no bias current is applied, there is no measurement. But
if the bias current slightly below $I_{T0}$ is applied, a ``$J$"
measurement is performed.

The Hamiltonian for a flux qubit is~\cite{majer}
\begin{equation}
H=I_p(\Phi_{e}-\frac{1}{2}\Phi_0)\sigma_z+\Delta \sigma_x
\end{equation}
where $I_p$ is the maximum persistent super-current of the flux
qubit, $\Delta$ is the tunneling amplitude of the barrier and
$\Delta\ll I_p\Phi_0$, with $\Phi_0$ being the flux quantum.
Initially we can set $\Phi_{e}\ll\Phi_0/2$ so that the state
$|00\rangle$ is produced for the two flux qubits. We then shift
$\Phi_e$ to $\Phi_0/2$ very fast and apply $I_b$ frequently. After a
time period of $\pi/(2\sqrt 2\Delta)$, the entangled state
$|\psi^+\rangle$ is produced if $V=0$ is verified throughout the
period. This procedure can be extended so as to experimentally
produce large cluster states. For existing technologies of
superconducting qubits, the detection time is around 1 {\em n}s,
while the decoherence time can be several {\em $\mu$}s (see, e.g.,
\cite{Bertet}), which indicates that thousands of $J-$measurements
could be done within the decoherence time. In the future, it would
be interesting to study the  effects of decoherence on this circuit.

\section{Concluding remarks}

We have studied the two-qubit quantum Zeno
effect with threshold detection, a type of non-deterministic
collective measurement: the $J-$measurement which distinguishes two
subspaces $J_1=\{|11\rangle\}$, and $J_0=\{|00\rangle, |01\rangle,
|10\rangle\}$. We show that the two-qubit quantum Zeno effect can be
used to produce and expand quantum entanglement, such as cluster
states, which are a useful resource for quantum computing. These give
new insights on the quantum Zeno effect and its possible application
in QIP.  The method presented here can also be used to produce other
types of entangled states, including the Greenberg-Horne-Zeilinger
states and the so-called ``$W$ states"~\cite{dur}. We also discussed
the possible implementation of the $J-$measurement with
superconducting qubits.

\section{acknowledgement}
F.N. gratefully acknowledges partial support from the National
Security Agency (NSA), Laboratory Physical Science (LPS), Army
Research Office (ARO), National Science Foundation (NSF) grant No.
EIA-0130383, JSPS-RFBR 06-02-91200, and Core-to-Core (CTC) program
supported by the Japan Society for Promotion of Science (JSPS).
X.B.W. was supported in part by the National Fundamental Research
Program of China grant Nos 2007CB807900 and 2007CB807901, NSFC grant
No. 60725416, and China High Tech. Program grant No. 2006AA01Z420.
J.Q.Y. was supported by the NSFC
grant Nos.~10534060 and 10625416, and the NFRPC grant No. 2006CB921205.

\end{document}